\documentclass[prl,aps]{revtex4}
\input epsf
\usepackage{graphics}
\usepackage{graphicx}

\begin{document}

\title{Backward Evolving Quantum States}

\author{Lev Vaidman%
\footnote{vaidman@post.tau.ac.il%
}}
\affiliation{ School of Physics and Astronomy\\
Raymond and Beverly Sackler Faculty of Exact Sciences \\
Tel-Aviv University, Tel-Aviv 69978, Israel}

\begin{abstract}
The basic concept of the two-state vector formalism, which is the
time symmetric approach to quantum mechanics, is the backward
evolving quantum state. However, due to the time asymmetry of the
memory's arrow of time, the possible ways to manipulate a backward
evolving quantum state differ from those for a standard, forward
evolving quantum state. The similarities and the differences between
forward and backward evolving quantum states regarding the
no-cloning theorem, nonlocal measurements, and teleportation are
discussed. The results are relevant not only in the framework of the
two-state vector formalism, but also in the framework of
retrodictive quantum theory.
\end{abstract}
\maketitle


\section{Introduction}

In 1964 Aharonov, Bergmann, and Lebowitz (ABL) \cite{ABL} laid the
foundations for the Time Symmetric Formalism of Quantum Mechanics.
The basic idea of the formalism is the introduction of a quantum
state evolving backward in time in addition to the standard quantum
state evolving forward in time. The  formalism uncovered numerous
peculiar features of quantum mechanics.

One peculiarity is that some particular pre- and post-selections
lead to situations in which we know with certainty the result of
some intermediate measurement, even though neither pre-selection nor
post-selection are eigenstates of the corresponding variable. I have
named such definite outcomes ``elements of reality'' \cite{ER}. It
is possible for pre- and post-selected systems to yield simultaneous
elements of reality for noncommuting observables \cite{XYZ}. Even
more surprisingly, it is possible to obtain seemingly contradictory
elements of reality, such as a single ball which is to be found with
certainty, if searched for, in any one of two or more distinct
locations \cite{AV1991}.

The most important result of the two-state vector formalism is
related to ``weak measurements'' performed on pre-and post-selected
systems. The outcomes of weak measurements (named {\it weak
values}), which are given by a simple universal expression, may turn
out to be far from the range of eigenvalues of the corresponding
operator \cite{AV1990}.

The backward evolving state is a premise not only of the two-state
vector formalism, but also of ``retrodictive'' quantum mechanics
\cite{retro}, which deals with the analysis of quantum systems based
on a quantum measurement performed in the future relative to the
time in question.

In this paper I shall not discuss the advantages of the two-state
vector and the retrodictive approaches. Rather, I consider here a
theoretical question: what are the limitations on the possible
manipulations of a backward evolving quantum state? In the following
sections I will analyze, in particular, the possibility of cloning
and teleporting backward evolving quantum states. I will touch upon
these questions also with regard to two-state vectors, which include
both forward and backward evolving states.

First, however, I would like to take advantage of the opportunity to
give tribute to GianCarlo Ghirardi and to express my view on the
Quantum Universe. In the next section I will explain how this view
is related to the backward evolving quantum state.

\section{The Quantum Universe}

There is no consensus regarding the Quantum Universe. Without having
seen other contributions, I take the risk of predicting that the
multitude of views presented here will show this very vividly.

The followers of Bohr forbid us to talk about the Quantum Universe:
quantum theory is not to tell us the story of ``reality''. I
believe, however, that science must tell us what the Universe is as
well as how it behaves. If quantum theory does not describe our
world, then what does? No other theory agrees so well with
experimental evidence.

For those who do wish to understand the Universe - and to accept
that it is quantum - the central question becomes: what is the
status, or physical meaning, of the quantum state, of the quantum
wave function? The Wave Function of the Universe, evolving according
to the Schrodinger Equation with its numerous branches, seems very
remote from the picture of the world as we experience it.

This connection is made via the idea that every measurement causes a
{\it collapse} of the wave function. For, the collapsed wave
function does resemble the world we know. When and how collapse
occurs, however, is a very difficult issue. GianCarlo Ghirardi,
together with Phillip Pearle, Riminy and Weber are the pioneers of
physical proposals for the collapse process \cite{P,GRW}.

In the past, David Albert and I analyzed the Stern-Gerlach
experiment within the framework of the Ghirardi-Rimini-Weber (GRW)
collapse model \cite{AVGRW}. We claimed that GRW-collapse does not
occur during a carefully arranged measurement process, but Ghirardi
and his collaborators showed that it does take place in the brain of
the observer \cite{AVGRWrep}. At that time David Albert persuaded me
of the advantages of the many-worlds interpretation, although he has
since come to favor the GRW type solution \cite{Al}.

Another option is to deny that the collapse process exists and that
the Schrodinger Wave corresponds to our experience. Bohm, de
Broglie, and Bell suggested that the Schrodinger Wave is just a {\it
pilot wave} of ``Bohmian particle positions'' and that those
``particles'' which have single trajectories draw the picture of our
familiar world.

My strong preference remains to deny both the collapse process and
the existence of additional ontological entities (i.e. Bohmian
particle positions). I accept that the Wave Function of the Universe
is all the physical ontology that exists. For this economy and
elegance in physical laws I am prepared to pay the price of
acknowledging that I and the world I see are not the only ones of
the physical universe. There are numerous parallel worlds similar to
the one I know, all incorporated in a single quantum State of the
Universe \cite{myMWI}. The parallel worlds are essentially
irrelevant to me in {\it this} world. In theory, parallel worlds can
interfere, but that this should be observable is not feasible.

I can trace back and understand my world in the past, but my (this
branch) future consists of multiple worlds. For the purpose of
describing a quantum system between two measurements it is
convenient to consider a single world defined by the results of
measurements both in the past and in the future relative the time I
wish to discuss. Thus, I find it convenient to use the two-state
vector formalism in the framework of the many-worlds interpretation.
The concept of the two-state vector can be understood most clearly
in a world with definite results of experiments both in the past and
in the future.

\section{Backward Evolving Quantum State}

There is an ongoing discussion about the meaning of the quantum
state. I have also taken part in this discussion \cite{psi}, but
here I shall adopt a pragmatic approach. A quantum state represents
a set of statements about the probabilities for the outcomes of
measurements that can be explained in the framework of any
interpretation of quantum mechanics. Let us spell out these
probabilities for backward and forward evolving quantum states.

We consider a quantum system which is described completely by a
variable $A$. At time $t$, an ideal measurement of $A$ is performed
with the outcome $A=a$. Then, it is said, immediately after the
measurement the system is described by the quantum state
$|A=a\rangle$ (evolving forward in time). This outcome also tells us
that immediately before the measurement the quantum state (evolving
forward in time) was not orthogonal to $|A=a\rangle$.

It is possible also to apply retrodiction to the measurement of $A$
and say that immediately prior to the measurement there is an
additional quantum state evolving backward in time $\langle A=a|$,
see Fig. 1.


\begin{figure}[h]
\includegraphics[width=7.0cm]
{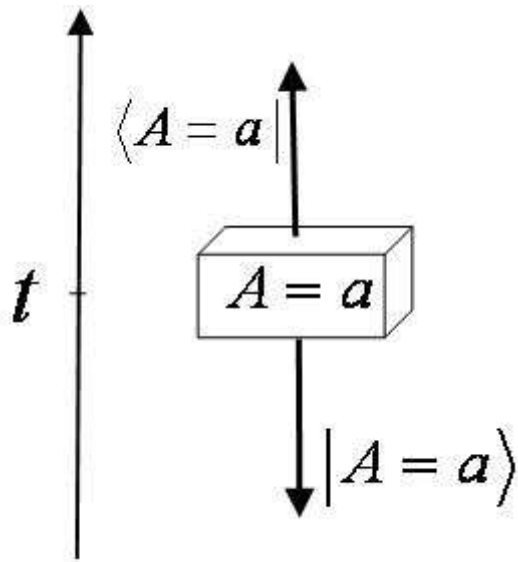}
  \caption{Measurement of $A$ at time $t$ creates
  the forward evolving quantum state $|A=a\rangle$ and the backward evolving
   quantum state $\langle A=a|$. } \label{1}
\end{figure}

The operational meaning of the fact that a system is described at a
particular time by the quantum state $|A=a\rangle$ evolving forward
in time, is that a measurement of $A$ at that time yields $A=a$ with
probability 1. This is irrespective of a post-selection of a
particular outcome of a measurement performed after that time. (Note
that cases in which another measurement of $A$ yields a different
outcome are automatically disregarded, since a quantum system in the
state $|A=a\rangle$, with the only interaction being another
measurement of $A$, cannot yield any other outcome in the latter
measurement of $A$.) Given that there is no post-selection, we can
state also that the probability of obtaining $B=b$ in a measurement
of another variable $B$ is given by the square of the scalar product
$|\langle B=b|A=a\rangle|^2$.

Almost the same can be said about the operational meaning of
describing a system at a particular time by the backward evolving
quantum state $\langle A=a|$. A measurement of $A$ at that time
yields $A=a$ with probability 1, irrespective of a pre-selection of
a particular outcome of a measurement before that time. (Note that,
again, cases in which we pre-select states orthogonal to
$|A=a\rangle$ are automatically disregarded.) If there is no
pre-selection, the probability of finding $B=b$ is given by the
square of the scalar product $|\langle B=b|A=a\rangle|^2$.

Usually, at the ``present'' time there is no particular
post-selection, but there is some pre-selection. In order to
eliminate this pre-selection we have to ``erase'' the past of the
quantum system. One way to achieve this is to cause the system to
become completely entangled with another ancillary system, see Fig.
2. So long as the ancilla is not measured, the future of the ancilla
is unknown and the past of the quantum system becomes unknown too.
 Thus it is possible for the
probabilities of the results of a measurement at a particular time
(after preparation of the entangled state and before the
post-selection) to depend solely on the backward evolving state.

\begin{figure}
  \includegraphics[width=12cm]{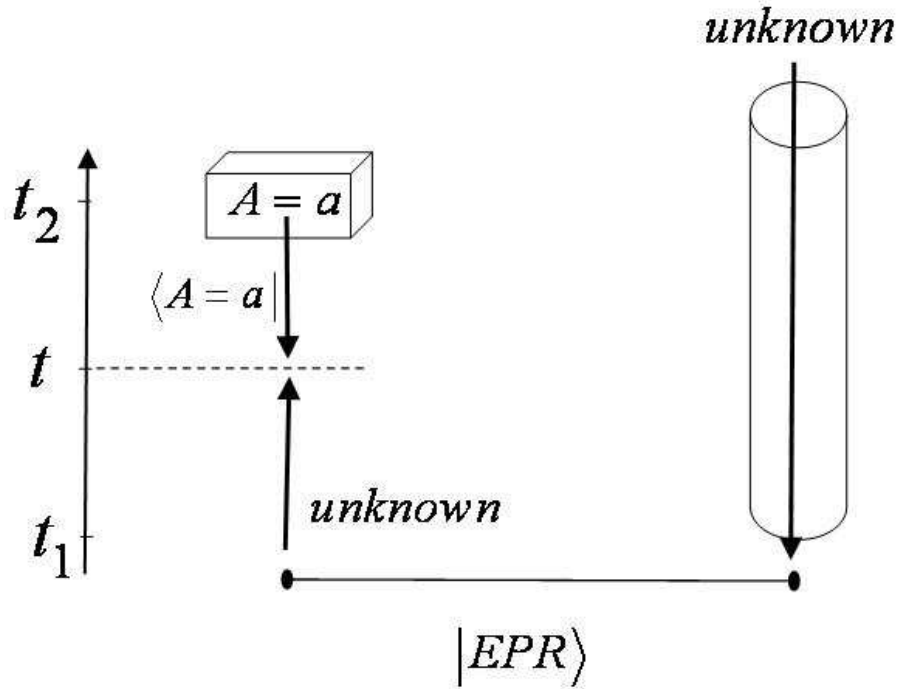}\\
  \caption{Erasure of the past of the particle with
   a backward evolving quantum
state  $\langle A=a|$. We prepare the EPR state of the particle and
an ancilla and guard the ancilla such that it will not be measured.}
\end{figure}

A formal proof of the above claim might be helpful at this point.
Consider two systems pre-selected in a completely entangled state $|
\Psi _{ent} \rangle$. One of the systems is post-selected via a
complete measurement whose outcome is $A=a$. We are looking for the
probability of obtaining the particular outcome $B=b_n$ in an
intermediate measurement of $B$. The probability of obtaining a
particular outcome in a measurement performed on a pre- and
post-selected system is given by the Aharonov-Bergmann-Lebowitz
(ABL) formula \cite{ABL}. Its generalization for partial
post-selection \cite{V98} yields:

\begin{eqnarray}
\label{ABL-new} \nonumber
 {\rm Prob}(B=b_n) &=& {{ || {\bf P}_{A=a}
{\bf P}_{B=b_n} | \Psi \rangle_{ent} ||^2} \over{\sum_i||  {\bf
P}_{A=a} {\bf P}_{B=b_i} | \Psi _{ent}\rangle ||^2}}\\ \nonumber
 &=& {{ ||~
|A=a\rangle \langle A=a| |B=b_n\rangle \langle B=b_n| \sum_j
|B=b_j\rangle | j \rangle_{anc} ||^2} \over{\sum_i||~|A=a\rangle
\langle A=a| |B=b_i\rangle \langle B=b_i| \sum_j  |B=b_j\rangle | j
\rangle_{anc} ||^2}} \\ \nonumber &=&{{ ||~ |A=a\rangle\langle A=a|
B=b_n\rangle  | n \rangle_{anc} ||^2} \over{\sum_i||~|A=a\rangle
\langle A=a| B=b_i\rangle | i \rangle_{anc} ||^2}} \\&=&{{ |~
\langle A=a| B=b_n\rangle |^2} \over{\sum_i|  \langle A=a|
B=b_i\rangle  |^2}}=|~ \langle A=a |B=b_n\rangle|^2.
\end{eqnarray}

This shows the equivalence, in some sense, of a quantum state at
present evolving forward in time when the future is unknown, and a
quantum state evolving backward in time given that the past of the
system is erased. ``Erasing'' the past is, in principle, a
complicated procedure. It consists of preparing an entangled state
of our system and an ancillary system, as well as of ``guarding''
the ancillary system against measurement. In many practical
situations, however, nature provides the ``erasing'' for free. For
example, the filament of a light bulb produces photons with an
``erased'' (for all practical purposes) polarization state. The
photon polarization is entangled with the quantum state of the
filament which cannot be measured using current technology. Leggett
\cite{Leg} called this process by a different name, ``garbling
sequence'' or ``coherence-destroying manipulation'', and mentioned
that nature, in effect, does it for us all the time.

\section{The Asymmetry of the Memory Time Arrow}

We wish to analyze the similarities and the differences between
forward and backward evolving quantum states, with regard to the
possibilities for performing various manipulations. One similarity
which we have already mentioned is that an ideal complete
measurement at a particular time that yields $A=a$ creates both the
quantum state $|A=a\rangle$ evolving forward in time and the quantum
state $\langle A=a|$ evolving backward in time.

A notable difference between forward and backward evolving states
has to do with the creation of a particular quantum state at a
particular time. In order to create the quantum state $|A=a\rangle$
evolving forward in time, we measure $A$ before this time. We cannot
be sure to obtain $A=a$, but if we obtain a different result $A=a'$
we can always perform a unitary operation and thus create at time
$t$ the state $|A=a\rangle$. On the other hand, in order to create
the backward evolving quantum state $\langle A=a|$, we measure $A$
after time $t$. If we do not obtain the outcome $A=a$, we cannot
repair the situation, since the correcting transformation has to be
performed at a time when we do not yet know which correction is
required. Therefore, a backward evolving quantum state at a
particular time can be created only with some probability, while a
forward evolving quantum state can be created with certainty. (Only
if the forward evolving quantum state is identical to the backward
evolving state we want to create at time $t$, and only if we know
that no one touches the system at time $t$, can the backward
evolving state be created with certainty, since then the outcome
$A=a$ occurs with certainty. But this is not an interesting case.)

The formalism of quantum theory is time reversal invariant. It does
not have an intrinsic arrow of time. The difference with regard to
the creation of backward and forward evolving quantum state follows
from the ``memory's'' arrow of time. We can base our decision of
what to do at a particular time only on events in the past, since
future events are unknown to us. The memory time arrow is
responsible for the difference in our ability to manipulate forward
and backward evolving quantum states.

\section{ Ideal Nondemolition Measurements and Teleportation}

The same ideal (von Neumann) measurement procedure applies both to
forward evolving quantum states and to backward evolving quantum
states. In both cases, the outcome of the measurement is known after
the time of the measurement. All that is known about what can be
measured in an ideal (nondemolition) measurement of a forward
evolving quantum state can be applied also to a backward evolving
quantum state. The proof relies on the fact that the operational
meaning of measurability for a forward evolving state, and that of
measurability for a backward evolving state, are essentially the
same: three consecutive measurements have to yield the same outcome.

There are constraints on the measurability of nonlocal variables,
i.e. variables of composite systems with parts separated in space.
When we consider instantaneous nondemolition measurements (i.e.
measurements in which, in a particular Lorentz frame during an
arbitrarily short time, local records appear which, when taken
together, specify the eigenvalue of the nonlocal variable), we have
classes of measurable and unmeasurable variables. For example, the
Bell operator variable is measurable, while some other variables
\cite{VP}, including certain variables with product state
eigenstates \cite{NLWE,NLWEVG}, cannot be measured.

The procedure for measuring nonlocal variables involves entangled
ancillary particles and local measurements, and can get quite
complicated. Fortunately, there is no need to go into detail in
order to show the similarity of the results for forward and backward
evolving quantum states. The operational meaning of the statement
that a particular variable $A$ is measurable is that in a sequence
of three consecutive measurements of $A$ - the first taking a long
time and possibly including bringing separate parts of the system to
the same location and then returning them, the second being short
and nonlocal, and the third, like the first, consisting of bringing
together the parts of the system - all outcomes have to be the same,
see Fig. 3. But this is a time symmetric statement; if it is true,
it means that the variable $A$ is measurable both for forward and
backward evolving quantum states.

\begin{figure}
  \includegraphics[width=14cm]{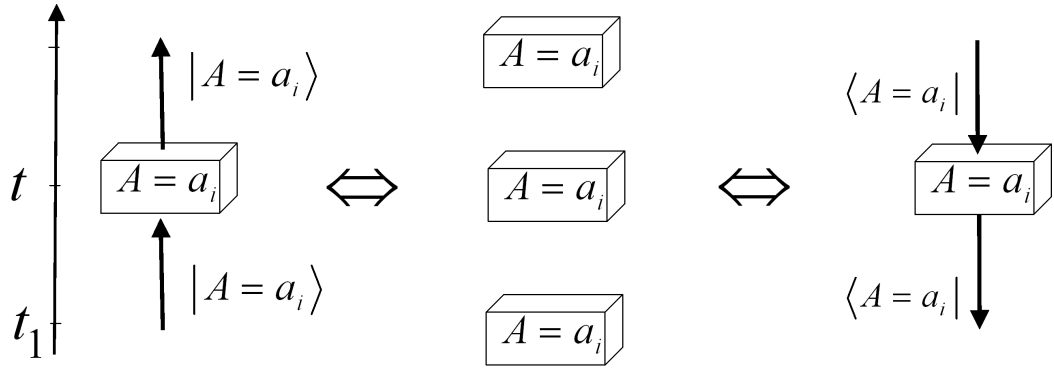}\\
    \caption{The operational meaning of creating a measuring device for a nondemolition measurement of
   $A$ at time $t$ means that the device and the two other measuring devices
   operating before and after time $t$ have to give the same eigenvalues of $A$.
   It is true both for forward and backward evolving quantum
   states.}
\end{figure}

We need also to obtain the correct probabilities in the case that
different variables are measured at different times. For a forward
evolving quantum state it follows directly from the linearity of
quantum mechanics. For a backward evolving quantum state, the
simplest argument is the consistency between the probability of the
final measurement, which is now $B=b$, given the result of the
intermediate measurement $A=a$, and the result of the intermediate
measurement given the result of the final measurement. We assume
that the past is erased. The expression for the former is $|~
\langle A=a |B=b\rangle|^2$. For consistency, the expression for the
latter must be the same, but this is what we need to prove.

In exactly the same way we can show that the same procedure for
teleportation of a forward evolving quantum state \cite{tele} yields
also teleportation of a backward evolving quantum state. As the
forward evolving quantum state is teleported to a space-time point
in the future light cone, the backward evolving quantum state is
teleported to a point in the backward light cone. Indeed, the
operational meaning of teleportation is that the outcome of a
measurement in one place is invariably equal to the outcome of the
same measurement in the other place. Thus, the procedure for
teleportation of the forward evolving state to a point in the future
light cone invariably yields teleportation of the backward evolving
quantum state to the backward light cone.

Let us consider an experiment demonstrating teleportation of a
backward evolving quantum state. Victoria performs a measurement of
$B$ and gives the quantum system to Alice. Alice teleports the
quantum state in the usual way to Bob who gives the state to Victor.
Victor measures $A$ and selects all cases in which he obtains $A=a$.
For these cases he asks Victoria for her measurement outcomes. The
outcomes should respect the probability law $p(B=b)=|\langle
B=b|A=a\rangle|^2$. In particular, when $B=A$, Victoria always (in
all the cases in which Victor asks her) gets $A=a$.

The impossibility of teleportation of the backward evolving quantum
state outside the backward light cone follows from the fact that it
will lead to teleportation of the forward evolving quantum state
outside the forward light cone, and this is impossible since it
obviously breaks causality.

\section{Demolition Measurements}

The argument used above does not answer the question of whether it
is possible to measure nonlocal variables in a demolition
measurement. Obviously, a demolition measurement of a nonlocal
variable of a quantum state evolving forward in time does not
measure this variable for a quantum state evolving backward in time.
Any nonlocal variable of a composite system can be measured with
demolition for a quantum state evolving forward in time \cite{V}.
Recently, it has been shown \cite{VN} also that any nonlocal
variable can be measured for a quantum state evolving backward in
time. Moreover, the procedure is  simpler and requires fewer
entanglement resources.

The difference follows from the fact that we can change the
direction of time evolution of a backward evolving state along with
flipping it, see Fig. 4. Indeed, all we need is to prepare an EPR
state of our system and an ancilla. Guarding the system and the
ancilla ensures that the forward evolving quantum state of the
ancilla is the flipped state of the system. For a spin wave function
we obtain
 $\alpha \langle{\uparrow}
|+\beta\langle{\downarrow}| \rightarrow
   -\beta^\ast |{\uparrow}\rangle +\alpha ^*|{\downarrow}\rangle $.
For a continuous variable wave function $\Psi (q)$ we need the
original EPR state $|q-\tilde{q}=0, ~~p+\tilde{p}=0\rangle$. Then,
the backward evolving quantum state of the particle will transform
into a complex conjugate state of the ancilla $ \Psi(q) \rightarrow
\Psi^\ast(\tilde{q})$.

\begin{figure}
 \includegraphics[width=14.5cm]{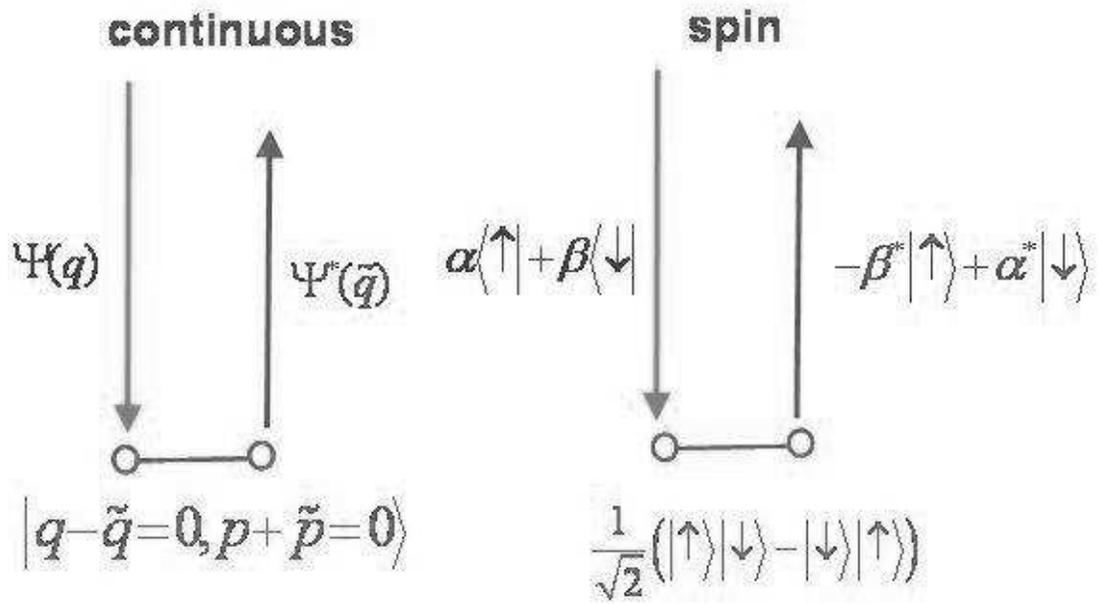}\\
  \label{6}
  \caption{The backward evolving state can be
transformed to ``flipped'' state evolving forward in time.}
\end{figure}

If the particle and the ancilla are located in different locations,
then such an operation is a combination of time reversal and
teleportation of a backward evolving quantum state of a continuous
variable \cite{V94}.

We cannot flip and change the direction of time evolution of a
quantum state evolving forward in time. To this end we would have to
perform a Bell measurement on the system and the ancilla and to get
a particular result (singlet). However, we cannot ensure this
outcome, nor can we correct the situation otherwise. Moreover, it is
easily proven that no other method will work either. If one could
have a machine which turns the time direction (and flips) a forward
evolving quantum state, then one could prepare at will any state
that evolves toward the past, thus signalling to the past and
contradicting causality.

\section{No Cloning Theorem}

The time symmetric approach to quantum mechanics suggests that the
no cloning theorem \cite{NC1,NC2} should be correct also for
backward evolving quantum states. However,  I do not see a
straightforward ``time reversed'' proof of the theorem. Due to the
memory time arrow which is not changed in our gedanken experiments,
the ``time reversed'' no cloning theorem for a backward evolving
state is different from the standard no cloning theorem. In the
normal case, our tools include measurements and interactions which
can be controlled by the results of the measurements. These
interactions take place after the measurement. In the time reversed
case, operations are to take place before the measurement. This
implies unusual capabilities, and the question becomes: can we clone
a quantum state when we have a machine that performs an operation on
our quantum system which is controlled by future measurement? I also
do not see how to adapt the standard proof of the no cloning theorem
based on linearity. The fact that various outcomes of the final
(post-selection) measurement have different probabilities presents a
serious obstacle.

\begin{figure}
 \includegraphics[width=14.5cm]{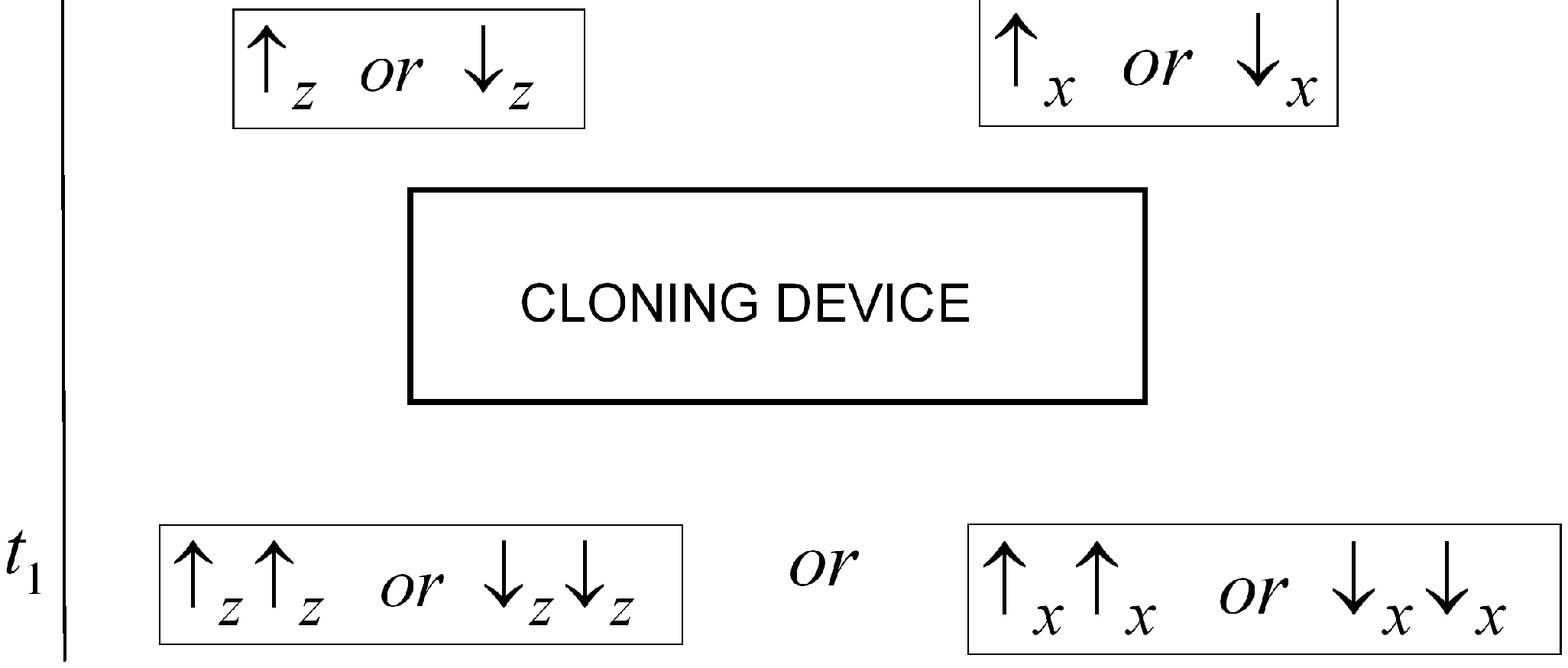}\\
\caption{Cloning device for backward evolving quantum state allows
sending signals to the past. Decision between measurement of
$\sigma_z$ or $\sigma_x$ at time $t_2$ leads to distinguishable
mixtures at the earlier time $t_1$.}
\end{figure}

The proof of the no cloning theorem which I have found relies on an
argument of causality. (Note the similarity to the history of the
discovery of the no cloning theorem \cite{NC2}). It is a
contradiction with causality to be able to send signals to the past.
Let us assume that we have a black box which performs cloning of
quantum states evolving backward in time. We arrange an ensemble of
pairs of, spin-$1\over 2$ particles. At time $t$, we measure the
spin in the $z$ direction of  particles in one half of the ensemble
of pairs, and the spin in the $x$ direction of  particles in the
other half of the ensemble. Then we operate our backward evolving
cloning machine which supposedly clones the state of one particle of
each pair to the other particle of the pair. Finally, we, in the
future, perform either a spin $z$ or a spin $x$ measurement on one
particle from each pair. Given that the cloning machine works well,
the pairs of the particles will be either in identical spin $z$
states or in identical spin $x$ states. Thus, at time $t$, either
all pairs of one half of the ensemble yield the same outcomes, or
all pairs of the other half yield the same outcomes, depending on
our decision of what to measure at a later time. The cloning machine
sends signals to the past. This proves that the cloning machine for
backward evolving quantum state does not exist.

\section{Manipulation of the Two-State Vector}

We have analyzed what can and what cannot be done for a backward
evolving quantum state. It turns out that every manipulation, except
``preparation'', which can be performed on the standard, forward
evolving quantum state, can be performed on the quantum state
evolving backward in time. Moreover, contrary to the case of the
quantum state evolving forward in time, it is possible to change the
direction of time evolution (together with flipping) of quantum
state evolving backward in time. This process can also be combined
with teleportation. Given a quantum channel (an entangled pair
separated in space) we can teleport and flip the direction of time
evolution without sending any classical information. This
``teleportation'' can be performed to any space-time point, and is
not constrained to the forward light cone as the standard
teleportation is.

Let us now briefly discuss  a quantum system described at a
particular time by the two-state vector. It is meaningless to ask
whether we can perform a nondemolition measurement on a system
described by a two-state vector. Indeed, the vector describing the
system should not be changed {\it after} the measurement, but there
is no such time: for a forward evolving state, ``after'' means
later, whereas for a backward evolving state, ``after'' means
before. It is meaningful to ask whether we can perform a {\it
demolition} measurement on a system described by a two-state vector.
The answer is positive \cite{VN}, even for composite systems with
separated parts.

Next, is it possible to teleport a two-state vector? Although we can
teleport both forward evolving and backward evolving quantum states,
we cannot teleport the two-state vector. The reason is that the
forward evolving state can be teleported only to the future light
cone, while the backward evolving state can be teleported only to
the backward light cone. Thus, there is no space-time point to which
both states can be teleported.

Finally, the answer to the question of whether it is possible to
clone a two-state vector is negative, since neither forward evolving
nor backward evolving quantum states can be cloned.

It is a pleasure to thank Tamar Ravon for helpful discussions. This
work has been supported by the European Commission under the
Integrated Project Qubit Applications (QAP) funded by the IST
directorate as Contract Number 015848.


\end {document}